\documentclass[prl,twocolumn,superscriptaddress,preprintnumbers,nofootinbib]{revtex4-1}
\usepackage[subpreambles=true]{standalone}

\usepackage{graphicx}
\usepackage{dcolumn}
\usepackage{bm}
\usepackage{hyperref}
\usepackage{subfigure}
\usepackage{float}
\usepackage{wrapfig}
\usepackage{amssymb,latexsym, amsmath}
\usepackage{graphics}
\usepackage{float}
\usepackage{appendix}
\usepackage{amssymb,latexsym,amsthm,makeidx,enumerate}
\usepackage{mathrsfs}
\usepackage{siunitx}
\usepackage{comment,hhline,multirow,setspace}

\usepackage{dcolumn}
\usepackage{bm}
\usepackage[T1]{fontenc}
\usepackage{amsmath}
\usepackage{graphicx}
\usepackage{tabularx}
\usepackage{mathtools}
\usepackage{xfrac}
\usepackage{orcidlink}
\usepackage{hyperref}\hypersetup{colorlinks=true, linkcolor=blue, citecolor=blue, urlcolor=blue, pdftitle={Direct observation of discommensurate charge density wave modulation in the quasi-1D Weyl semimetal candidate NbTe4}, pdfauthor={P. Giraldo-Gallo}}
\newcommand{\NT}{NbTe$_{4}$}
\newcommand{\TT}{TaTe$_{4}$}
\graphicspath{{graphics/}}

\begin{document}

\makeatletter
\let\NAT@bare@aux\NAT@bare
\def\NAT@bare#1(#2){%
	\begingroup\edef\x{\endgroup
		\unexpanded{\NAT@bare@aux#1}(\@firstofone#2)}\x}
\makeatother

\makeatletter
\newcommand{\printfnsymbol}[1]{%
  \textsuperscript{\@fnsymbol{#1}}%
}
\makeatother

\title{Direct observation of discommensurate charge density wave modulation in the quasi-1D Weyl semimetal candidate NbTe$_4$}
\author{J. A. Galvis\orcidlink{0000-0003-2440-8273}}
\affiliation{Faculty of Engineering and Basic Sciences, Universidad Central, Bogot\'a, Colombia}
\author{A. Fang\orcidlink{0000-0002-2634-4448}}
\affiliation{Department of Applied Physics and Geballe Laboratory for Advanced Materials, Stanford University, Stanford, California 94305, USA}
\affiliation{Stanford Institute for Materials and Energy Sciences, SLAC National Accelerator Laboratory, 2575 Sand Hill Road, Menlo Park, California 94025, USA}
\author{D. Jiménez-Guerrero}
\author{J. Rojas-Castillo}
\affiliation{Department of Physics, Universidad de Los Andes, Bogot\'a 111711, Colombia}
\author{J. Casas}
\author{O. Herrera}
\affiliation{Faculty of Engineering and Basic Sciences, Universidad Central, Bogot\'a, Colombia}
\author{A. C. Garcia-Castro\orcidlink{0000-0003-3379-4495}}
\affiliation{School of Physics, Universidad Industrial de Santander, Carrera 27 Calle 09, 680002, Bucaramanga, Colombia}
\author{E. Bousquet\orcidlink{0000-0002-9290-3463}}
\affiliation{Physique Th\'eorique des Mat\'eriaux, QMAT, CESAM, Universit\'e de Li\`ege, B-4000 Sart-Tilman, Belgium}
\author{I. R. Fisher\orcidlink{0000-0002-1278-7862}}
\author{A. Kapitulnik\orcidlink{0000-0002-2569-3582}}
\affiliation{Department of Applied Physics and Geballe Laboratory for Advanced Materials, Stanford University, Stanford, California 94305, USA}
\affiliation{Stanford Institute for Materials and Energy Sciences, SLAC National Accelerator Laboratory, 2575 Sand Hill Road, Menlo Park, California 94025, USA}
\author{P. Giraldo-Gallo$^{\dagger}$\orcidlink{0000-0002-2482-7112}}
\affiliation{Department of Physics, Universidad de Los Andes, Bogot\'a 111711, Colombia}

\date{\today}
\begin{abstract}

The transition-metal tetrachalcogenides are a model system to explore the conjunction of correlated electronic states such as charge density waves (CDW), with topological phases of matter. Understanding the connection between these phases requires a thorough understanding of the individual states, which for the case of the CDW in this system, is still missing. In this paper we combine phonon-structure calculations and scanning tunneling microscopy measurements of \NT{} in order to provide a full characterization of the CDW state. We find that, at short range, the superstructure formed by the CDW is fully commensurate with the lattice parameters. Moreover, our data reveals the presence of phase-slip domain-walls separating regions of commensurate-CDW in the nanoscale, indicating that the CDW in this compound is discommensurate at long-range. Our results solve a long-standing discussion about the nature of the CDW in these materials, and provide a strong basis for the study of the interplay between this state and other novel quantum electronic states.

\end{abstract}

\maketitle

\def\thefootnote{$\dagger$}\footnotetext{Corresponding author: pl.giraldo@uniandes.edu.co}\def\thefootnote{\arabic{footnote}}

\section*{Introduction}

Transition-metal chalcogenides are a model family to study a large variety of quantum phenomena including charge density wave (CDW) formation, superconductivity, topological states and magnetism \cite{Manzeli2017,Monceau2012,DELIMA2019}. Among this family, the quasi-1-dimensional transition-metal tetrachalcogenides, with their representative members: \NT{} and \TT{}, have recently gained a renewed interest due to the possibility of hosting axionic states connecting different Weyl-points in the Fermi-surface through the formation of a CDW \cite{Jian2020,Gooth2019,Sun_2020,Jian2020}. Therefore these compounds provide an ideal opportunity to study the connection between topological states of matter and correlated electronic states such as CDWs \cite{Boswell83,Walker85}. In addition, recent high-pressure experiments have reported the presence of superconductivity in both, \NT{} \cite{Yang2018} and \TT{} \cite{Yuan2020}, making them strong candidates for a new class of topological superconductors. All of these novel phenomena motivate a thorough characterization of the different physical properties of these materials, in particular, of the details and origin of the CDW at low-temperatures, where the possible topological properties are more likely to be observed.

\begin{figure}[!t]
\centering
\includegraphics[width=0.50 \textwidth]{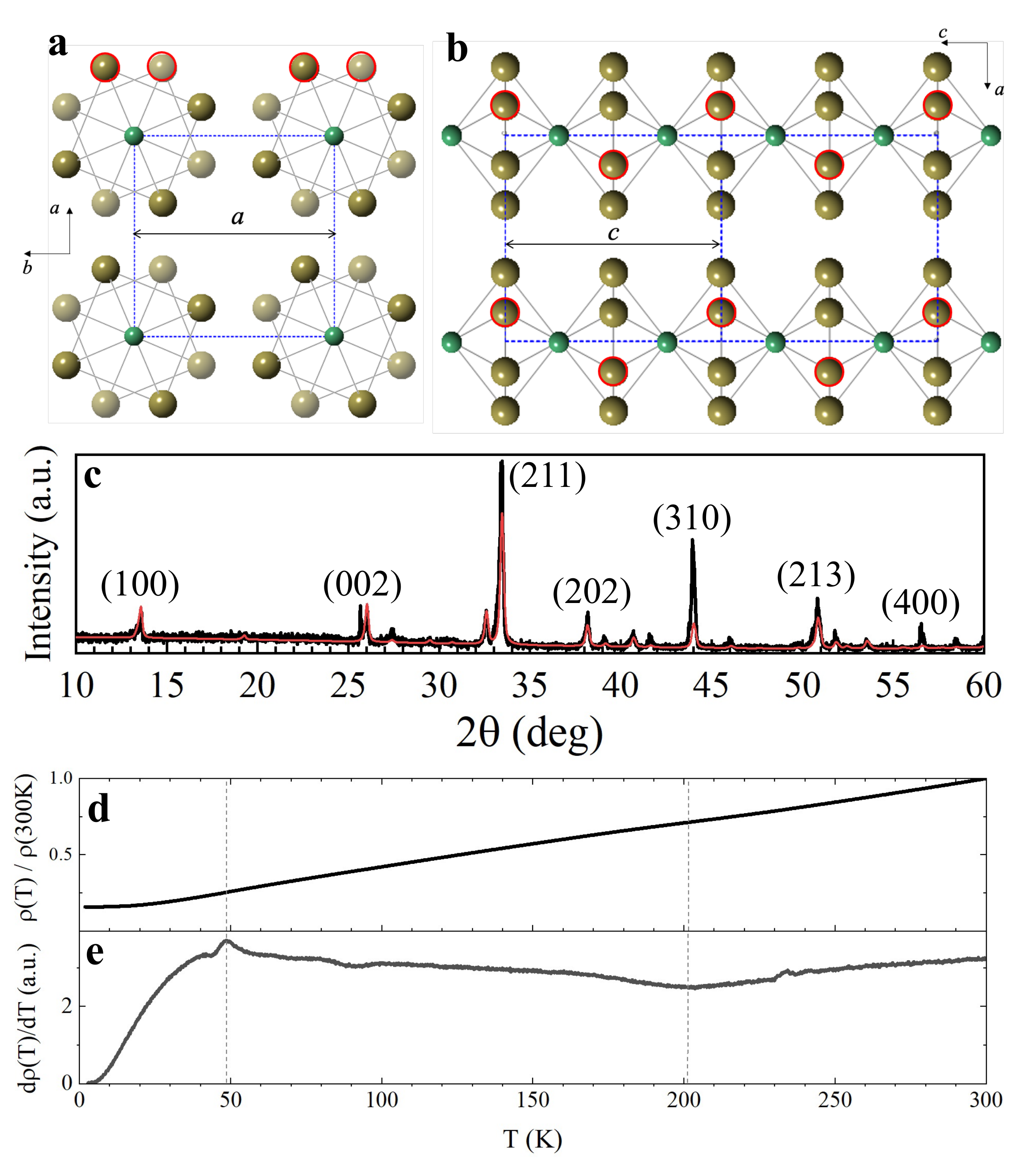}
\caption{(Color online) \textbf{(a)} Crystal structure of \NT{}, with space group \textbf{P4/mcc} (SG. 124) projected along the \textbf{(a)} [001] and \textbf{(b)} [010] directions. The Te-atoms with red borders in (a) and (b) are atoms forming a cleaved \{010\} surface. \textbf{(c)} Powder X-ray diffraction measurements  of ground single crystals (black-line), with diffraction peaks consistent with the \textbf{P4/mcc} space group (refinement in red).  \textbf{(d)} Temperature dependence of the resistivity, normalized by the room-temperature value. The residual resistivity ratio (RRR) is 6.3, similar to previously reported studies \cite{Yang2018,Ikari_1987,TADAKI1990}. \textbf{(e)} Derivative of the normalized resistivity curve in (d). Vertical-dashed-lines indicate changes in the temperature dependence of the resistivity, previously associated to CDW-transitions.
}
\label{fig_crystal}
\end{figure}

The crystal structure of \NT{} originates from subtle deformations of its high-temperature tetragonal space group \textit{P4/mcc}, shown in figure \ref{fig_crystal}(a,b), with reported unit cell parameters $a,b$=6.499 \AA{} and $c$=6.837 \AA{} at room-temperature, and confirmed by our x-ray data (fig. 1c) \cite{Selte1964}. In this structure, Nb-atoms are at the center of two square antiprisms of Te-atoms, each rotated with respect to the other (fig. \ref{fig_crystal}a). These units form linear chains in the \textit{c}-direction (fig. \ref{fig_crystal}b), which are bonded to other chains by van der Waals interactions. As a consequence, single crystals of this compound grow as long-needles parallel to the \textit{c}-axis, which evidences the quasi-one-dimensional character of its crystal structure.

X-ray and electron diffraction studies in \NT{} have revealed distortions from the \textit{P4/mcc} structure, and have associated those to several possible CDW distortions \cite{Walker88,Boswell83,Prodan87}. A variety of satellite peaks and streaks in the diffraction patterns develop at different temperatures and the appearance of such features roughly coincide with features observed in electrical resistivity measurements (see vertical-dashed-lines in fig. \ref{fig_crystal}(d,e) for some of such features) \cite{TADAKI1990}. At room-temperature, satellite peaks at $\bm{q_1}=(\frac{1}{2}a^*,\frac{1}{2}a^*,\left(\frac{2}{3}+\delta\right)c^*)$, $\bm{q_2}=(0,0,\left(\frac{2}{3}-2\delta\right)c^*)$, $\bm{q_3}=(0,0,\left(\frac{2}{3}+4\delta\right)c^*)$, where $a^*$ and $c^*$ are the inverse of $a$ and $c$, respectively, and $\delta=0.022$ (following the notation of \textit{Eaglesham et al.}) are observed by selected-area diffraction patterns as reported by several authors \cite{Eaglesham_1985, Boswell86, Prodan87}. Below approximately 210 K to 180 K (varies for different authors and on cooling/warming history), diffuse streaks at different positions of the reciprocal space are progressively developed, and are even reported to evolve into regularly spaced spots as temperature is decreased  \cite{Prodan87}. Below about 50 K, the different satellite diffraction peaks become commensurate with the \textit{P4/mcc} lattice parameters (meaning, $\delta=0$), and the $\bm{q_2}$ and $\bm{q_3}$ peaks merge into one ($\bm{q_2}=(0,0,\frac{2}{3}c^*)$). This has been interpreted as a lock-in (incommensurate-to-commensurate) transition of the CDW distortion \cite{Eaglesham_1985}.

Although the characterization of the lattice distortion and reciprocal space mapping of \NT{} by diffraction experiments has been extensive, direct observation of the CDW modulation by real-space imaging techniques is scarce. Previous scanning tunnelling microscopy (STM) data attempted to characterize the charge modulation at room-temperature, however, with a limited resolution \cite{Prodan98}. To our knowledge, there are no previous reports of the observation of the low-temperature CDW modulation of \NT{} by STM experiments. Recent experiment in the other family member of the tetrachalcogenides, \TT{}, reported on the observation of a new type of charge modulation, distinct from the ones reported by diffraction experiments, presumably coming from surface effects in this compound \cite{Sun_2020}. All of these observations motivate a thorough characterization of the low-temperature CDW states in \NT{} by direct imaging and spectroscopic techniques.

In this article we report on a combination of low-temperature STM measurements and phonon dispersion calculations of the quasi-1D transition-metal tetrachalcogenide \NT{}. Our measurements of the low-temperature charge distribution in the $\{010\}$ plane directly reveal, for the first time, the superstructure created by the CDW in this material, which agrees with our theory predictions. Furthermore, we present evidence for the presence of a discommensurate-CDW (i.e., commensurate regions separated by phase-slip domain-walls) at low-temperature, with wavevector and phase-slip conditions consistent with the features reported by previous diffraction experiments \textit{at room-temperature}, although our measurements are performed at 1.7 K. This observation suggests that the CDW in this compound is discommensurate for all temperatures below room-temperature, therefore inviting to revisit the characteristics and conditions of the presumable lock-in transition and other intermediate-temperature transitions reported for this material. Our results provide strong evidence to solve a long-standing discussion around the characteristics of the CDW in this material, and provides a strong basis for the study of the interplay between this state and other novel quantum electronic states, such as the recently predicted Weyl-points.

\begin{figure}[!t]
\centering
\includegraphics[width=0.48 \textwidth]{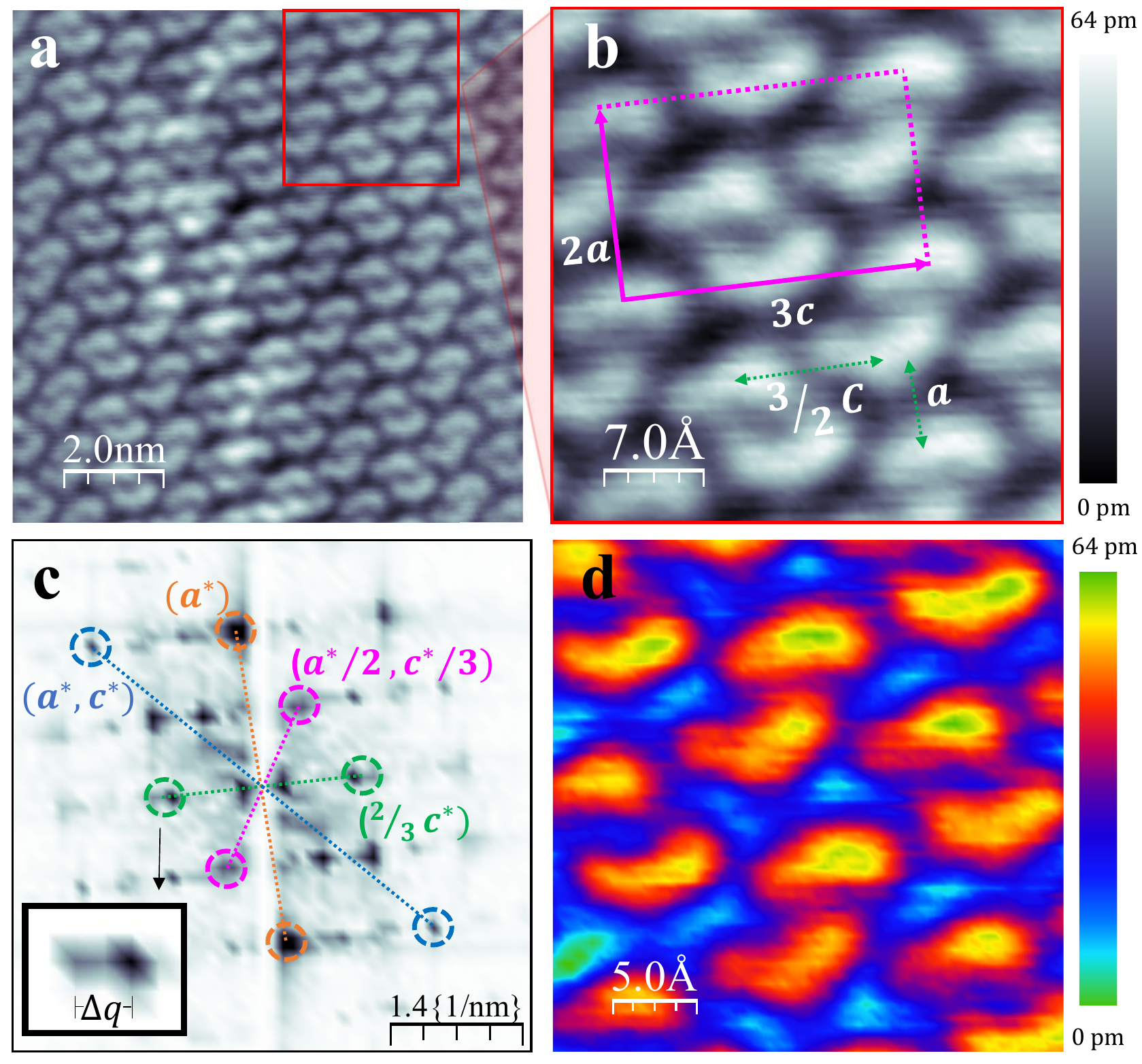}
\caption{\textbf{(a)} Scanning tunneling topography image of a cleaved NbTe$_{4}$ single crystal. The CDW superstructure formed at the Te-terminated surface is observed. The area enclosed by the red square is magnified in \textbf{(b)} in which the unit cell of the superstructure (pink square), with size 2$a\times$3$c$, can be clearly identified. \textbf{(c)} Fourier transform of the image shown in (a). Different peaks have been highlighted by different-color circles. The peaks highlighted by the pink-circles correspond to the CDW superstructure unit cell. The peaks highlighted by the green-circles correspond to the distance between adjacent ``cashews'' (Te-trimers) with opposite curvature, along the $c$-direction, this is, with central $q=(0,\frac{2}{3}c^*)$. Theses peaks are split, as better seen in the inset. \textbf{(d)} Close-up look to a STM image, presented using a color scale that highlights the asymmetry in the charge distribution of the cashews or Te trimers. The maximum of the scale (yellow-colors) is shifted toward the right-side for all cashews, independent of their orientation (facing up or down).
}
\label{fig_STM}
\end{figure}

\section*{Methods}
\subsection{Computational Approach}
The density functional theory, DFT \cite{PhysRev.136.B864, PhysRev.140.A1133}, calculations were performed with the \textsc{abinit} package (v8.11.8) \cite{ABINIT, gonze2020}. 
The norm-conserving pseudopotentials from the PseudoDojo project (v0.4) \cite{Setten2018} were used with the generalized gradient approximation PBEsol exchange correlation functional \cite{perdew2008} and a Fermi-Dirac electronic smearing of 0.001 Hartree (Ha). 
The phonon calculations were performed using the density functional perturbation theory (DFPT) \cite{gonze1997} and a good convergence (about 1 cm$^{-1}$ on the frequencies) was obtained with a $k$-points mesh of $8\times8\times8$ in the reciprocal space and a cutoff energy for the plane wave expansion of 35 Ha. 
The phonon dispersions were interpolated with a grid of $4\times4\times4$ $q$-points in the reciprocal space. 
The symmetry adapted mode contributions to the phase-transition was analyzed using the \textsc{AMPLIMODE} software \cite{Orobengoa:ks5225}.

\subsection{Experimental Methods}
Single crystals of NbTe$_4$ were grown using a self-flux technique \cite{2012Fisher}. A mixture of 1 mol\% of elemental Nb and 99 mol\% of elemental Te were put in alumina crucibles and sealed in evacuated quartz tubes. The mix was heated to 700$^{\circ}$C, held at this temperature for 12 hours and then slowly cooled to 500$^{\circ}$C at a rate of 2$^{\circ}$C/hour. The remaining melt was decanted and separated from the NbTe$_4$ crystals using a centrifuge. Silver-colored long rectangular prism shaped crystals were obtained, with sizes up to 0.1 $\times$ 0.1 $\times$ 1 cm$^3$.

Powder X-ray diffraction measurements were taken from a collection of ground single crystals, in Eulerian-Cradle geometry, using a Panalytical X-Pert system. Resistivity measurements as a function of temperature (Fig. 1d,e) were taken in cleaved crystals using a four probe configuration, with electrical current running along the \textit{c}-direction. Electrical contact was made using sputtered gold pads. A $^4$He VTI-cryostat was used to vary temperature from room-temperature down to 2 K.

STM was performed with a hybrid UNISOKU-USM1300 system constructed with a homemade ultrahigh-vacuum sample preparation and manipulation system. The samples were cleaved at room-temperature at pressures lower than $10^{- 10}$ torr and immediately transferred to the low temperature STM. Crystals cleave exposing the \{010\} surface, this is, the \textit{a-c} plane. Topography was performed at 1.7 K, with typical tunneling parameters of $V_{bias}$=50–100 mV and $I$=100–300 pA.

\section*{Results and discussion}

Figure \ref{fig_STM} shows STM measurements taken at the \{010\} surface of a NbTe$_{4}$ crystal. The cleaved-surface that is observed in the image is formed by Te-atoms in the a-c plane, as the bonding between Te-atoms is weaker than between Nb and Te \cite{Selte1964}, a common feature in all low-dimensional transition-metal chalcogenides. The Te-atoms expected to be observed in the surface are highlighted in red-circles in Fig. \ref{fig_crystal}b, forming zig-zag chains that run along the \textit{c}-direction. The large-scale topographic image (figure \ref{fig_STM}a) reveals a periodic modulation of the electronic density, formed by rows of cashew-shaped units with alternating up-down curvatures, and with adjacent rows shifted 1/3 of a row period. This modulation forms a superstructure with unit cell as highlighted by the pink-rectangle in the close-up look to the topographic image, shown in fig. \ref{fig_STM}b. The spatial periodicity can be accurately determined through the fast Fourier transform (FFT) in Fig. \ref{fig_STM}c, resulting in a $2a \times 3c$ periodicity (FFT peaks labeled as $(a^*/2,c^*/3)$). The size of this superstructure matches the commensurate-CDW superstructure reported by several diffraction experiments at low-temperatures ($<$50 K) \cite{Boswell83,Eaglesham_1985,Boswell86}. We will later discuss other important features of the FFT of the topographic image. 

\begin{figure}[!t]
\centering
\includegraphics[width=0.5 \textwidth]{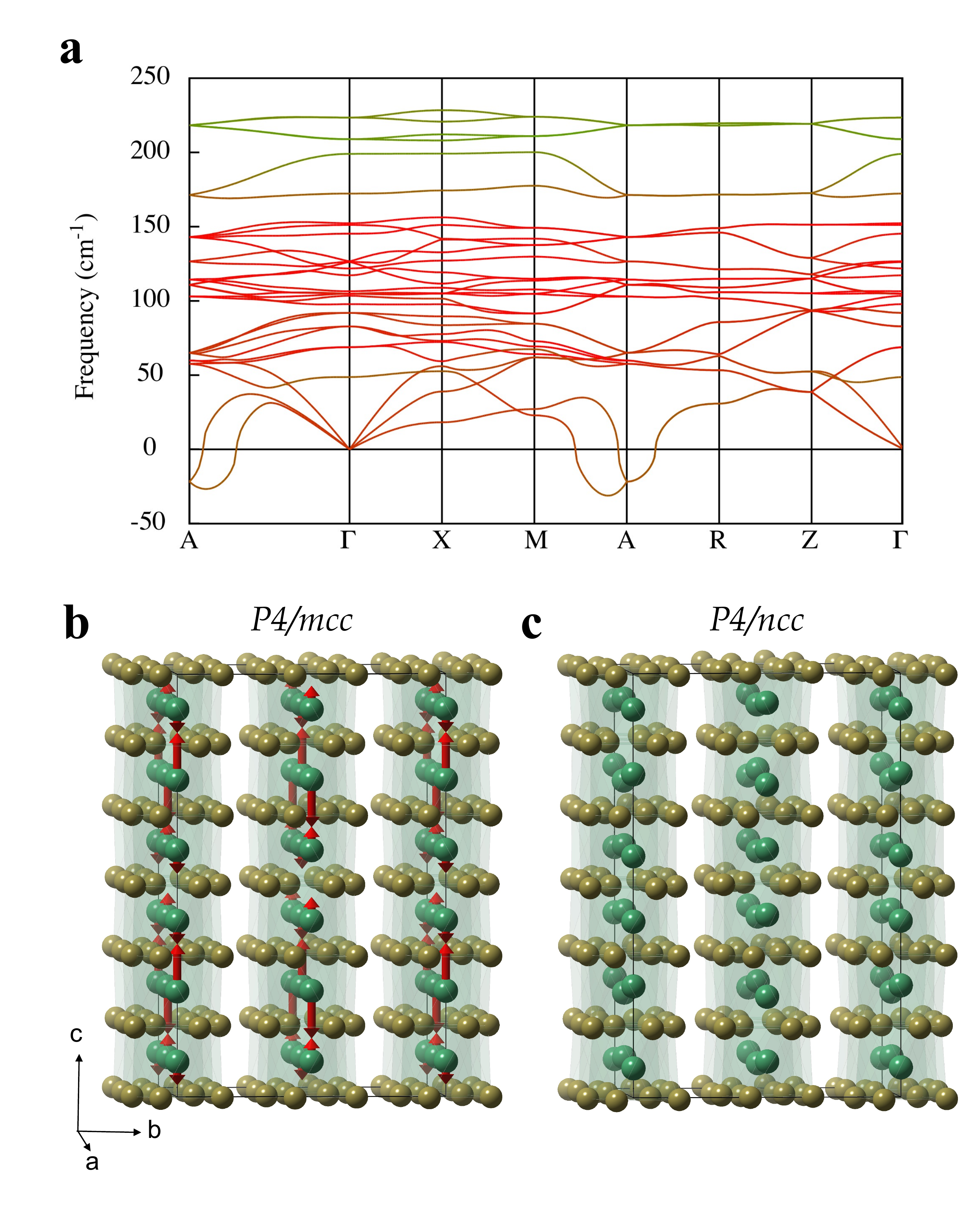}
\caption{(Color online) \textbf{(a)} Calculated phonon dispersion curves along various paths between high-symmetry points in the \emph{P4/mcc} Brillouin zone of NbTe$_4$. The unstable modes with imaginary frequencies are presented by negative values on the plot. The dispersion line color has been assigned to each point according to the contribution of each kind of atom to the associated eigenvector (red for Nb and blue for Te). \textbf{(b)} Schematic picture of the eigendisplacements of the $V_4$ unstable phonon at the $q$=($\sfrac{1}{2}$,$\sfrac{1}{2}$,$\sfrac{1}{3}$) vector, the atom displacements are shown as red-arrows in the hihg-symmetry \emph{P4/mcc} phase. \textbf{(c)} Fully relaxed modulated \emph{P4/ncc} structure is presented.}
\label{fig_DFT}
\end{figure}

\begin{figure}[!t]
\centering
\includegraphics[width=0.5 \textwidth]{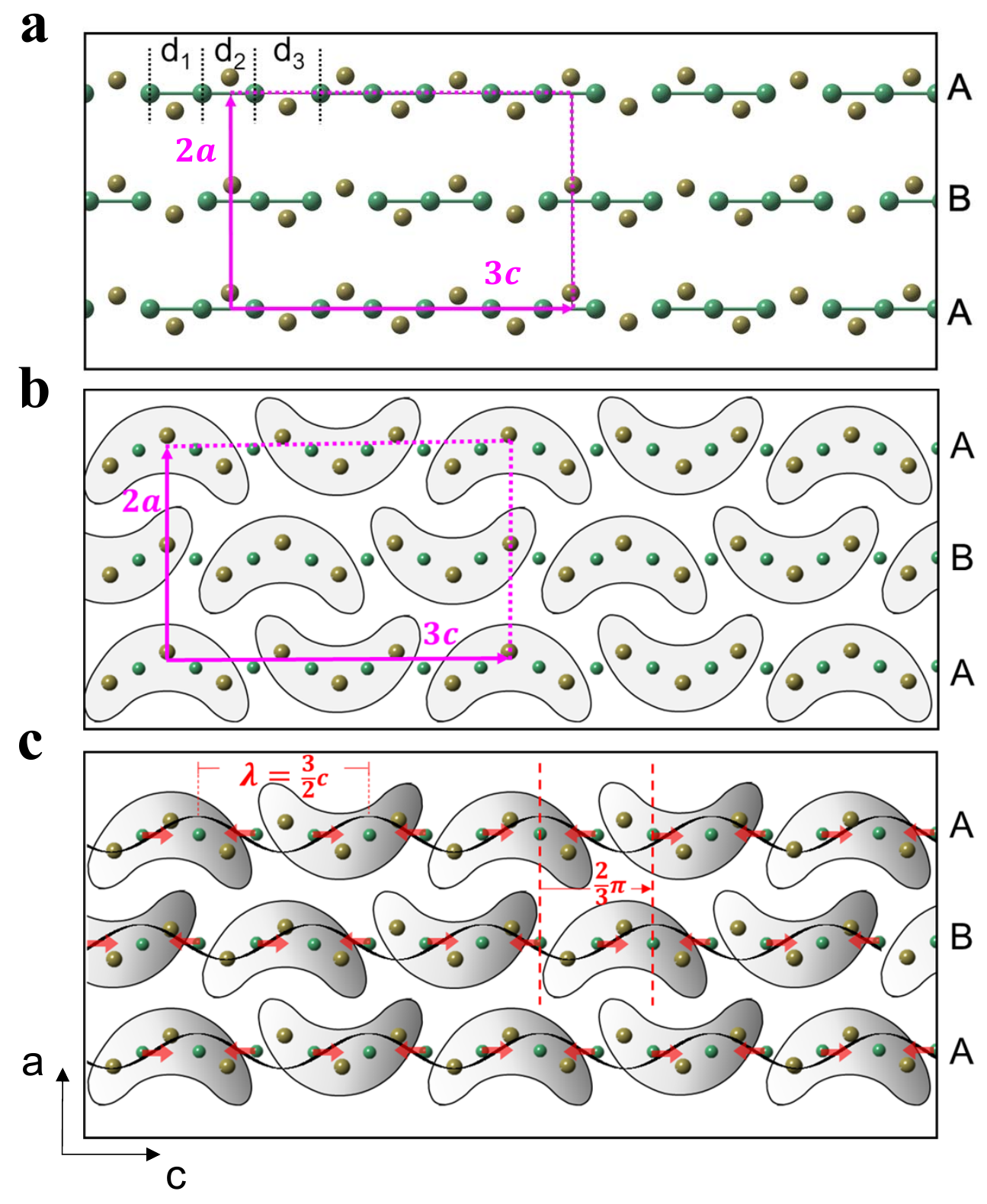}
\caption{(Color online) All figures show the Te-atoms (dark-yellow balls) of three chains in a $\{010\}$ cleaved-surface, as well as the closest Nb-atoms (green balls) underneath this surface. \textbf{(a)} Spatial configuration of the Nb-trimers in the final modulated structure as calculated by our DFT model. Three nonequivalent distances between the Nb-atoms are found in the model, with two short ($d_1$ and $d_2$) and one long ($d_3$), therefore forming Nb-trimers as represented by the green solid-lines. The unit cell of the Nb-trimers superstructure is indicated by the pink rectangle. A-B-A refers to the stacking of the chains. \textbf{(b)} Representation of the Te-trimer formation, which leads to the cashew-shapes observed in the topographic images. The unit cell of the Te-trimers superstructure is indicated by the pink rectangle, matching the unit cell of the Nb-trimers superstructure. \textbf{(c)} Red-arrows in this figure indicate a simplified representation of the Nb-atoms displacements, which lead to the CDW modulation in this compound. The charge density is maximum around the center of each Nb-trimer, which coincides with the right-side of the cashews, generating their asymmetric charge density (gradient color-fill). The charge density amplitude variation is depicted by the black-waves. The wavelength of the CDW is $\lambda=3c/2$, and the phase-shift between adjacent chains is $\pm 2\pi/3$. 
}
\label{fig_model}
\end{figure}

In order to explain the superstructure shown by the topography images and the origin of the CDW modulation in NbTe$_4$, we have computed the vibrational landscape of the high-symmetry structure \emph{P4/mcc} (space group (SG) 124).
Fig. \ref{fig_DFT}a shows the calculated phonon-dispersion curves of the \emph{P4/mcc} high-symmetry structure. An instability appears at the $A$ ($\sfrac{1}{2}$,$\sfrac{1}{2}$,$\sfrac{1}{2}$) high-symmetry point of the Brillouin zone (with the irreducible representation -- irrep -- $A_2$), which disperses quickly when going away from the $A$ point. 
However, the strongest unstable phonon frequency is located in a point between the $A$ and $M$ point, in a region close to $q$=($\sfrac{1}{2}$,$\sfrac{1}{2}$,$\sfrac{1}{3}$) (irrep $V_4$). To resolve the lowest energy structure, we condensed the $A_2$ and $V_4$ unstable mode eigenvectors into two different structures, with 2$a$$\times$2$a$$\times$2$c$ and 2$a$$\times$2$a$$\times$3$c$ supercells with respect to the high-symmetry unit cell, respectively. We find that the phase formed by the $V_4$ mode (which leads to the \emph{P4/ncc} - SG 130) has a lower energy than the phase formed by the $A_2$ mode (\emph{I4} - SG 79). This result is in good agreement with the observed 2$a\times$3$c$ superstructure seen at low-temperatures in our STM data.
Interestingly, our calculations show that the $V_4$-mode is strongly driven by the Nb-sites with a small deviation, from the high-symmetry position, of the Te-sites. Fig. \ref{fig_DFT}b shows the undistorted \emph{P4/mcc} phase in which the most significant atomic displacements of the $V_4$-mode are depicted by red-arrows. 
Once these distortions are condensed and the structure fully relaxed, the obtained structural modulation can be observed in Fig. \ref{fig_DFT}c.
In such a phase, the modulation induces a trimerization of the Nb-sites in the 1D-chains. The three Nb$-$Nb distances, $d_{Nb-Nb}$, are 3.137  \r{A}, 3.168 \r{A} and 3.971 \r{A}, hence, two short and one long. This result is in line with suggestions from previous works \cite{Walker88,2021Guster}. 
Fig. \ref{fig_model}a shows the spatial configuration of the Nb-trimers that lie just underneath a cleaved-surface, in the final modulated structure, as calculated by our DFT model. This figure highlights the A-B-A stacking of the trimers in adjacent chains, with the B-chain trimers shifted by one Nb-atom to the right (or two Nb-atoms to the left), or equivalently, $c/2$ to the right (or $c$ to the left), with respect to the A-chain trimers. This specific A-B-A stacking of the chains is responsible for the $2a$ periodicity of the superstructure along the direction perpendicular to the chains.

The trimerization of Nb-atoms, predicted by our DFT calculations, has important consequences for the charge distribution in the cleaved-surface of the \NT{} crystals. As mentioned previously, this surface is formed by Te-atoms. The schematic configurations shown in fig. \ref{fig_model}a,b,c shows these surface Te-atoms in a \{010\} cleaved-surface, together with the Nb-atoms just underneath this surface. The cashew-shapes observed in our topographic images can be naturally explained by the trimerization of the Te-atoms in the surface, which are arranged in zig-zag along the chains, as represented in fig. \ref{fig_model}b  (see figs. A1 and A2, in Appendix, for images with atomic-resolution within a cashew). The formation of Te-trimers (or cashews) is a consequence of the trimerization of the Nb-atoms underneath. This trimerization also naturally explains the alternating up-down curvature of successive cashews within a chain. The cashew configuration along each chain, with period $3c$, is shifted by $c$ in both -up and down- adjacent chains (to the left if taken with respect to the middle-B-chain in fig. \ref{fig_model}b, or equivalently, $2c$ to the right), and in the same direction for both adjacent chains. This specific A-B-A stacking of the chains is fully equivalent to the one found in our DFT calculations, and reproduces the unit cell of the superstructure, with period $2a\times 3c$, observed by our topographic images.

Noteworthy, the charge distribution within a single cashew-like structure is asymmetric, as is better appreciated in fig. \ref{fig_STM}d. The position where the maximum charge distribution occurs, as observed in the topography image, is shifted toward the right side on each cashew. This asymmetry is an indication that the CDW distortion indeed originates in the Nb-atoms, instead of the Te-atoms, in full agreement with our DFT calculations. 
Fig. \ref{fig_model}c shows a simplified schematic representation of the Nb-atoms displacements that give rise to such an asymmetric Te-surface charge distribution. In this picture, the central Nb-atom of each Nb-trimer, where charge density is expected to be higher, is located toward the right-end of the Te-trimer, which could explain the accumulation of charge toward the right-side of the cashews in the observed Te-surface. Within this picture, the CDW is represented by the waves depicted in Fig. \ref{fig_model}c, with a maximum amplitude right at (or very close to) the center of the Nb-trimer (which coincides with the right side of each Te-trimer or cashew), and a minimum amplitude in the space between two Nb-trimers. The wavelength of this CDW is $\lambda=3c/2$, and it therefore implies a CDW-wavevector within each chain of $q_{CDW}=(0,0,2c^*/3)$. Waves in both -up and down - adjacent chains of each chain are shifted by $2\pi/3$ either both to the left or both to the right, generating the observed A-B-A stacking of the superstructure. As a consequence of the CDW-wavector and the phase-shifts in adjacent chains, the observed superstructure in the Te-atoms-surface has a periodicity of $2a\times 3c$, as revealed by our topographic images, and predicted by our DFT calculations. Therefore, the phonon-mode responsible for such a commensurate distortion is the $V_4$-mode. The CDW-gap opened by this distortion, as determined by tunneling conductance curves, is $\Delta_{CDW} \approx$ 24 meV (see Fig. A3 in Appendix).

\begin{figure*}[t]
\centering
\includegraphics[width=1 \textwidth]{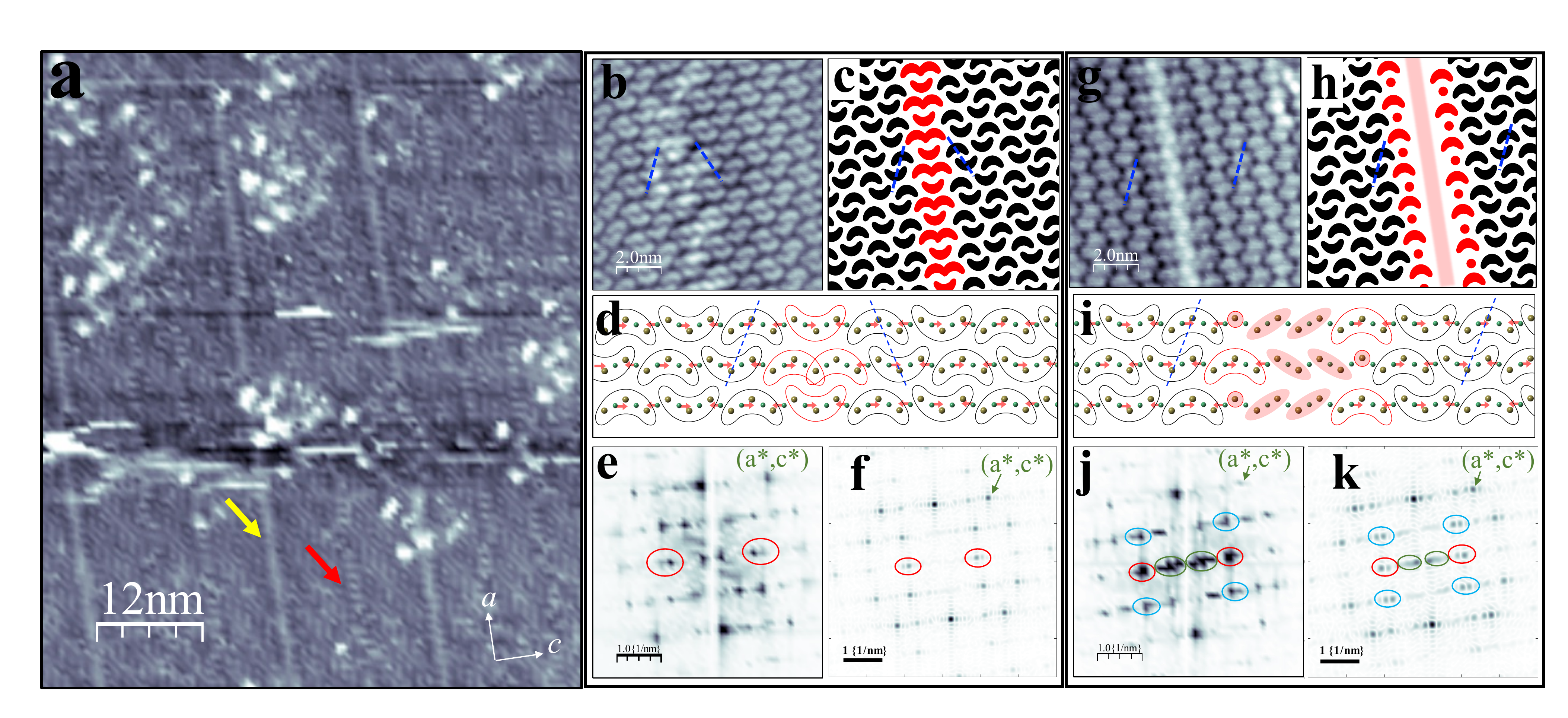}
\caption{(Color online) \textbf{(a)} Wide-area STM topographic image of a cleaved NbTe$_{4}$ single-crystal showing different types of domain-walls which run over long distances along the \textit{a}-direction. The yellow- and red-arrows highlight two different types of domain-walls, which generate different contrasts in the topographic image. \textbf{(b)} and \textbf{(g)} show topographic images of two types of domain-wall regions, and \textbf{(c)} and \textbf{(h)} show schematic representations of the cashews configuration across the domain-walls in (b) and (g), respectively. The cashews forming the domain-walls are represented in red. The blue-dashed-lines represent the sign of the phase-shift of the CDWs of two adjacent chains: a change in inclination represents a change of phase sign. The schematic representations in (c) and (h) capture the change of sign of this phase-difference at the left-side and right-side of the domain-walls. \textbf{(d)} and \textbf{(i)} show the atomic configuration of the Te-atoms at the surface (and the Nb-atoms underneath) which lead to the domain-walls of (b,c) and (g,h), respectively. \textbf{(e,f,j,k)} show the FFT of the images in (b,c,g,h), respectively. Red-ovals in all FFTs highlight the $(0,2c^*/3)$ peaks, which splitting is reproduced in the real-images FFTs and the schematic representation FFTs. Blue-ovals in the FFTs of (j,k) highlight the $(a^*/2,2c^*/3)$ peaks, which are split for the second type of domain-wall (g-k), but not for the first type (b-f). Green-ovals in the FFTs of (j,k) highlight peaks along the horizontal-axis and close to the origin, which are present for the second type of domain-wall, but not for the first type.
}
\label{fig_PS}
\end{figure*}


We will now discuss the intricacies of the FFT, which reveal the presence of discommesurations of the CDW in \NT{}. The FFT (fig. \ref{fig_STM}c) of the topographic image of fig. \ref{fig_STM}a reveals other intense peaks, in addition to the $(a^*/2, c^*/3)$ peaks of the superstructure unit cell. Interestingly, the peak around $(0, 2c^*/3)$ (highlighted in the green-circle) is split in two, as can be better observed in the inset to this figure. 
Although the presence of split peaks as well as the presence of other satellite peaks in diffraction experiments have been interpreted as signatures of incommensurate CDWs \cite{Boswell83,Prodan87,Prodan90}, we argue that all satellite peaks observed by diffraction experiments in all ranges of temperatures are a consequence of discommensurations of a unique commensurate CDW-phase, this is, regions of commensurate CDW separated by domain-walls. Figure \ref{fig_PS}a shows a wide-area topographic image of a cleaved \NT{} surface. Long columnar features along the \textit{a}-direction, and running from top-to-down all through the image, can be observed. An example of one of the features with the largest contrast is indicated by the yellow-arrow. However, other types of columnar features that generate less contrast in the topographic image can be observed, as for example, the one highlighted by the red-arrow. Careful and closer inspection to different areas containing such columnar features reveal different patterns. Two examples of the type of features that can be found are shown in Figs. \ref{fig_PS}b (same as in fig. \ref{fig_STM}a) and fig. \ref{fig_PS}g. The FFT of these images (figs. \ref{fig_PS}e and \ref{fig_PS}j, respectively) show marked differences. For instance, whereas the $(0,2c^*/3)$ peaks (red-ovals) in fig. \ref{fig_PS}e, for the first type of domain-wall, are split, as discussed earlier, the $(a^*/2,2c^*/3)$ peaks are not. In contrast, both of those set of peaks in the FFT shown in fig. \ref{fig_PS}j (red-ovals in this figure), for the second type of domain-wall, are split and, furthermore, additional peaks along the horizontal-axis, close to the origin (green-ovals), appear in this FFT. Fig. \ref{fig_PS}c shows a cartoon model of the distortion of the cashews pattern of the image in fig. \ref{fig_PS}b. The domain-wall is highlighted by the cashews in red. This type of domain-wall can be generated by the situation depicted in fig. \ref{fig_PS}d in which, within a chain, two nearest-neighbors Nb-trimers share one atom, implying a CDW phase-slip of $2\pi/3$ to the left within the chain. This, for the Te-trimers at the surface, results in the partial-merge of two cashews and therefore, the vertical flip of one of them. The Nb (and Te) trimers in adjacent chains remain intact, creating a domain-wall with alternating double-simple cashews. One way to validate this picture is observing the flip in CDW phase-difference between successive chains on the left and right of the domain-wall: with respect to the top-chain in fig. \ref{fig_PS}d, the CDW-phase of the middle row is shifted $2\pi/3$ to the left, on the left of the domain-wall, but  $2\pi/3$ to the right, on the right of the domain-wall. This is represented by the flip in orientation of the dashed-blue-lines in figs. \ref{fig_PS}(b,c,d). The FFT of the cartoon model in fig. \ref{fig_PS}c is shown in fig. \ref{fig_PS}f, and it reproduces the main features observed in the FFT of the corresponding real topographic image. 
The main characteristics of the second example of possible domain-walls, shown in fig. \ref{fig_PS}g, are captured by the cartoon model in fig. \ref{fig_PS}h. This pattern can be obtained by the situation shown in fig. \ref{fig_PS}i, in which five Nb-atoms within a chain are not clearly bound to trimers, deriving in a CDW phase-slip of $4\pi/3$ to the right, or equivalently, $2\pi/3$ to the left, within each chain. This phase-slip happens for every chain in the structure, which implies an equivalent CDW interchain phase-shift at the right and left of the domain-wall (situation highlighted by the dashed-blue-lines in figs. \ref{fig_PS}(g,h,i)), in contrast with the first example of domain-wall. The FFT of the cartoon model is shown in fig. \ref{fig_PS}k, and also reproduces the main features of the FFT of the corresponding real topographic image.

Other types of domain-walls, different from the two examples discussed previously, can be found within our topographic images. In principle, a domain-wall can be created by any possible combination of phase-slips for different chains. Nevertheless, only intrachain CDW phase-slips of $\pm 2\pi/3$ (or equivalently, $\pm n*c/2$ displacements of cashews, where $n$ is an integer) are allowed by the symmetry of the crystal structure and the CDW-wavelength, which limits the possibilities for domain-wall configurations. This constrain in the phase-slip can be the reason for the finite set of satellite peaks found in diffraction experiments. Moreover, the values for splittings reported by such diffraction experiments match the values found for splittings in the FFT peaks of our topographic images. The distance between the split peaks around the $(0, 2c^*/3)$ position in figs. \ref{fig_STM}c and \ref{fig_STM}e is $\Delta q=0.1975$ nm$^{-1}=0.132c^*$, which matches the value of the splitting between satellite diffraction peaks $\bm{q_2}=(0,0,2c^*/3-2\delta c^*)$ and $\bm{q_3}=(0,0,2c^*/3+4\delta c^*)$, i.e., $6\delta c^*$, with $\delta=0.022$. These values are equivalent to the ones reported by diffraction experiments by several authors \cite{Eaglesham_1985, Boswell83, Boswell86, Prodan87}. This highly suggests that, what had been interpreted before as signatures of the presence of incommesurate CDWs in \NT{} are, on the other hand, signatures of discommensurations due to phase-slip domain-walls, which separate regions of commensurate-CDW with wavevector $q_{CDW}=(0,0,2c^*/3)$ within a chain, phase-difference of $\pm 2\pi/3$ with adjacent chains, and ABA stacking. 

An important characteristics of the domain-walls is that they run over large distances along the same line in the \textit{a}-direction, with no (or minor) disruptions, and are not necessarily periodic along the \textit{c}-direction. The highly coherent domain-walls along the \textit{a}-direction may be a consequence of the interchain coupling strength, which for \NT{} is much more significant than for other low-dimensional transition-metal chalcogenides. For instance, the interchain Te-Te distance in the tetrachalcogenides is shorter than the intrachain Te-Te distance, which has been predicted to result in the formation of interchain Te-Te dimers that can hybridize with the transition-metal atoms, leading to a large metallic conduction in the direction perpendicular to the chains \cite{Bullett_1984,2021Guster}. In fact, the resistivity anisotropy of \NT{} has been reported to be very close to 1 \cite{TADAKI1990}, unexpected for such an anisotropic quasi-1D crystalline structure. The results presented by our study inspire relevant question that can help understanding these electronic properties. For example, the large tunnel conductance values observed in some of the domain-walls, which could be an indication of a largest charge density, invites to study the domain-walls contribution to the electronic conductance in the \textit{a}-direction. On the other hand, similarly long and coherent columnar domain-walls have been observed in other low-dimensional systems (for example, graphene nanostructures \cite{Vicarelli2015}), in which  topological defects due to atomic vacancies in the crystalline structure form extended dislocations. A similar mechanism could also be the origin of the observed CDW domain-walls in \NT{}, possibly originated by Te-vacancies.  Another relevant question is the influence of random quenched disorder on the domain-wall density, coherence length, and even CDW fundamental properties, as observed for other prototypical CDW systems \cite{2019Fang}.

It is important to emphasize that our STM measurements have been performed at low-temperature, and not at room-temperature. The presence of split peaks in diffraction experiments have been reported at room-temperature, with the appearance of other streaks and peaks as temperature is lowered. The equivalent characteristics of the FFTs of our topographic images and the ones reported by diffraction experiments suggest that the CDW in \NT{} is discommensurate and with an unchanged CDW-wavevector from room-temperature down to 1.7 K. Nevertheless, resistivity measurements as a function of temperature (see fig. 1d,e) reveal marked changes of slope at certain temperatures (both in our data and data taken by other authors), which have been historically interpreted as CDW phase-transitions. Interestingly, a temperature of $\sim$50 K, at which a marked feature in the derivative of the $\rho$ vs T curve appears (see fig. 1e), seems to coincide with the temperature below which the peak splitting and presence of diffuse streaks disappear in electron diffraction experiments, which has been interpreted as a lock-in (incommensurate-to-commensurate) transition of the CDW. As argued, the features observed in diffraction experiments could be explained by discommensurations, rather than incommensurations of the CDW, inviting to consider the configuration and dynamics of the phase-slip domain-walls as possible candidates driving the presumed phase-transitions suggested by the features in resistivity curves and diffraction experiments. 

Finally, it is pertinent to highlight that our topographic data can be fully reproduced by the condensed $V_4$ phonon-instability predicted by our DFT calculations. This indicates that an electron-phonon coupling mechanism is largely responsible for the commensurate-CDW formation in this compound, in contrast to the Fermi-surface nesting or Peierls mechanism commonly expected in low-dimensional compounds \cite{1988Gruner}. This observation is in line with previous works suggesting the electron-phonon mechanism as the dominant driving force for CDW formation in most real materials, even in the 1D limit \cite{2008Johannes,2021Guster}. Since the transition-metal tetrachalcogenides have been predicted to host Weyl-nodes in their electronic structure, our results motivate further theoretical and experimental studies aimed at understanding the influence of electron-phonon coupling on the determination of complex Fermi-surface topologies and its connection with novel topological quasiparticles in materials \cite{2013Parente,Heid2017,2021Garcia}.

\section*{Acknowledgements}
 The authors thanks C. A. Mera, L. Quiroga, F. Rodriguez, H. Suderow, I. Guillamon and E. Herrera for enlightening discussions. J.A.G., D.J-G., J.R-C., J.C., O.H. and P.G-G. thank the support of the Ministerio de Ciencia, Tecnología e Innovación de Colombia (Grants No. 120480863414 and No. 122585271058). P.G-G. and I.R.F. thank the support of the APS International Travel Grant Award Program which enabled P.G-G. to visit Stanford University. P.G-G. thanks the support of the L'Oréal-UNESCO For Women in Science International Rising Talents Programme, the School of Sciences and the Vice Presidency of Research and Creation at Universidad de Los Andes. J.A.G., J.C. and O.H. thank the support of Cl\'uster de Investigaci\'on en Ciencias y Tecnolog\'ias Convergentes NBIC, Universidad Central. I.R.F., A.F. and A.K. were supported by the Department of Energy, Office of Basic Energy Sciences, under contract DE-AC02-76SF00515. 
 
 This work used the DECI PRACE project OFFSPRING and the CECI facilities funded by F.R.S-FNRS (Grant No. 2.5020.1) and Tier-1 supercomputer of the F\'ed\'eration Wallonie-Bruxelles funded by the Walloon Region (Grant No. 1117545).
 
A.C.G.C. acknowledge the support from the GridUIS-2 experimental testbed, developed under the Universidad Industrial de Santander (SC3-UIS) High Performance and Scientific Computing Centre, with support from UIS Vicerrector\'ia de Investigaci\'on y Extensi\'on (VIE-UIS) and several UIS research groups as well as other funding resources.

	
\bibliography{1_bib_NbTe4.bib}

\newpage

\appendix
\section*{Appendix}
\setcounter{figure}{0} 
\makeatletter 
\renewcommand{\thefigure}{A\@arabic\c@figure}
\makeatother


\begin{figure}[b!]
\centering
\includegraphics[width=0.50 \textwidth]{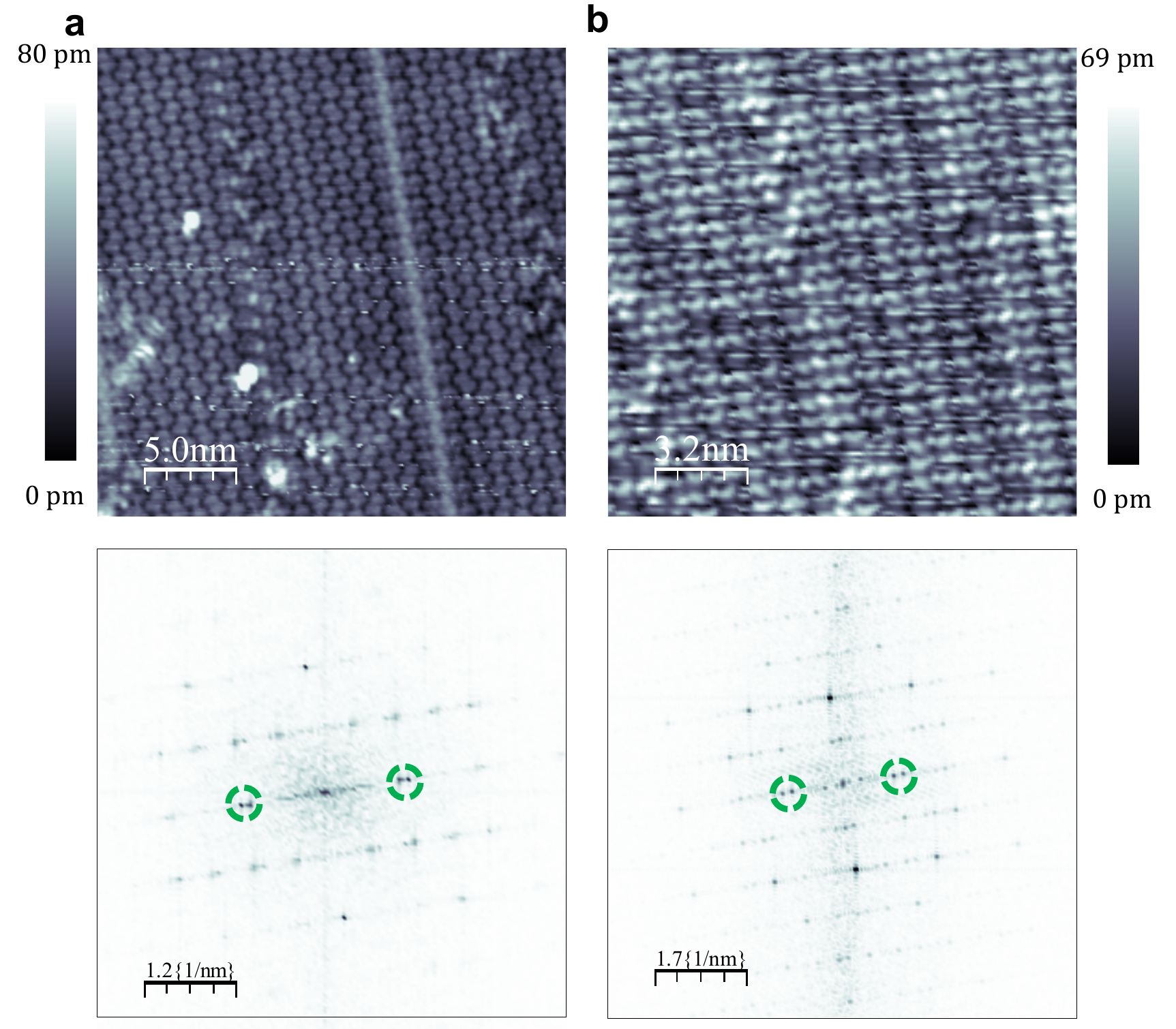}
\caption{(Color online) Scanning tunneling atomic resolution topographic images (top row) of NbTe$_4$. The image in \textbf{(b)} has enough resolution to identify the atomic structure of the cashew units. The bottom row shows the FFTs of the corresponding images on top. The $(0,2c^*/3)$ peaks are highlighted in green circles.
}
\label{fig_S1}
\end{figure}

\begin{figure}[b!]
\centering
\includegraphics[width=0.50 \textwidth]{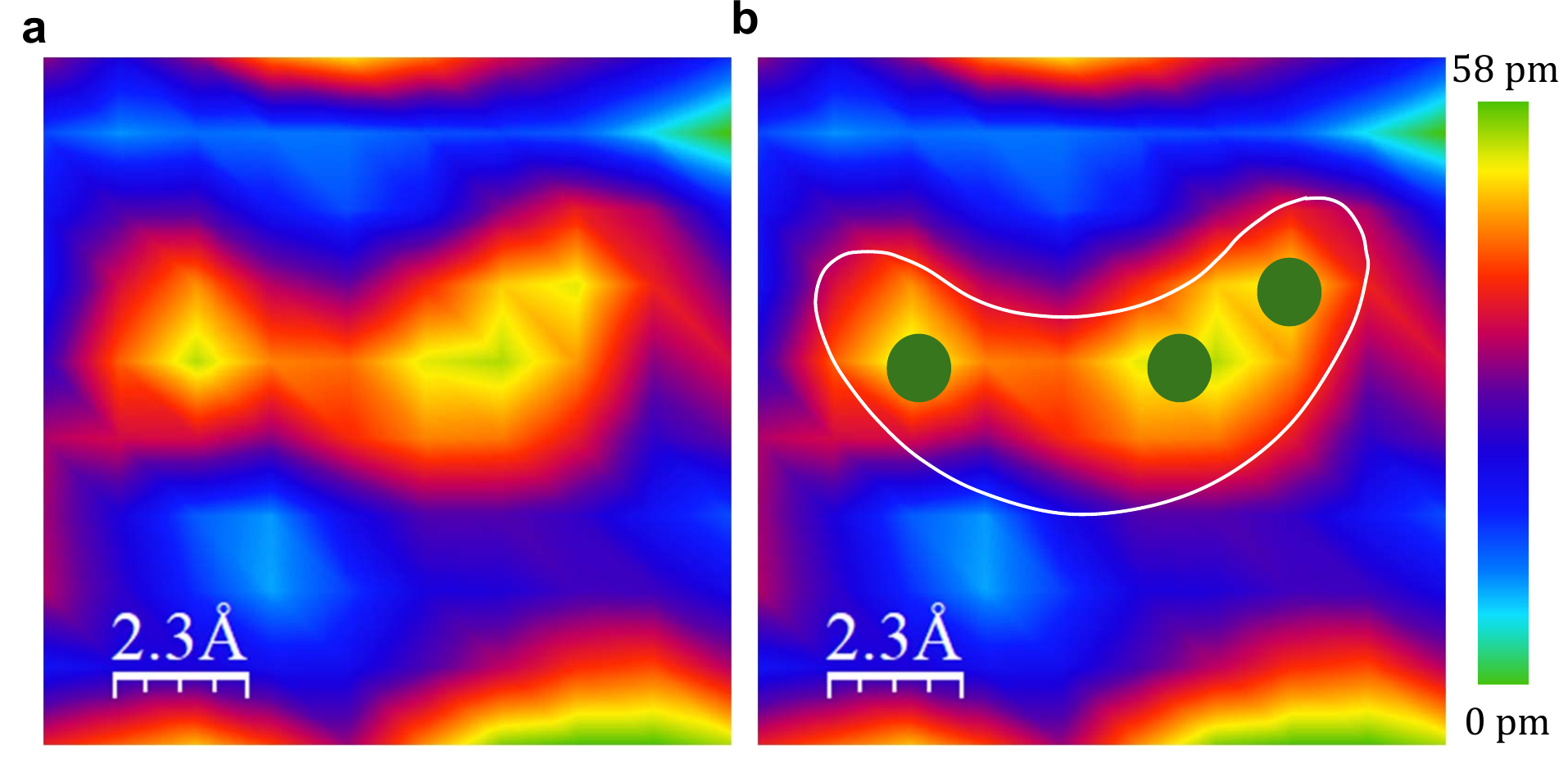}
\caption{(Color online) \textbf{(a)} Close-up look to a single cashew-unit in fig. \ref{fig_S1}(b), using a color scale that highlights the structure of each cashew. \textbf{(b)} Same as in (a), but with superimposed representation of the cashew shape and possible Te-atoms positions.
}
\label{fig_S2}
\end{figure}

Figure \ref{fig_S1} shows two scanning tunneling microscopy images in the \{010\} surface of NbTe$_4$, with different types of columnar domain walls in the CDW superstructure. The FFT of theses images (bottom panels of \ref{fig_S1}) show the splitting on the $(0,2c^*/3)$ peaks. The $(a^*/2,2c^*/3)$ peaks are split for the image in figure \ref{fig_S1}(a) but not for the image in figure \ref{fig_S1}(b). Interestingly, the STM image in \ref{fig_S1}(b) has a resolution enough for identifying individual Te atoms within each cashew. Fig \ref{fig_S2} shows the internal structure of a single cashew or Te-trimer in fig. \ref{fig_S1}(b). The three nonequivalent distances between the Nb atoms found in our theoretical model (fig 4 of the main text), give rise to an equivalent distortion on the Te atoms in the surface, as shown by figure \ref{fig_S2}b, leading to an asymmetric charge distribution within each cashew.       

\begin{figure}[t]
\centering
\includegraphics[width=0.40 \textwidth]{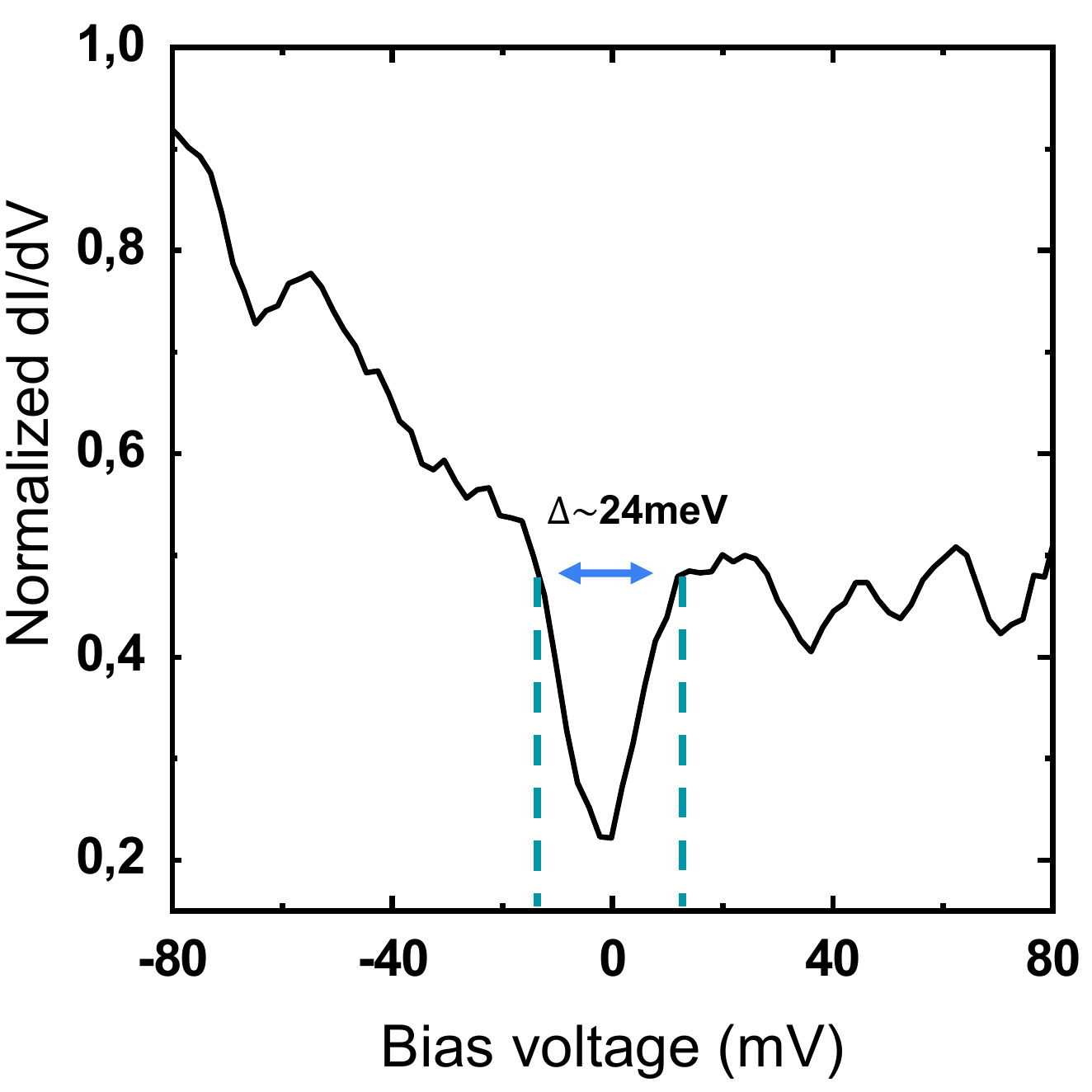}
\caption{(Color online) Tunneling conductance vs bias voltage curve taken at T = 1.7 K. The CDW gap measured is $\Delta_{CDW} \approx$ 24 meV.
}
\label{fig_S3}
\end{figure}

Figure \ref{fig_S3} shows a representative tunneling conductance vs bias voltage curve taken at a temperature of 1.7 K. The CDW gap measured is $\Delta_{CDW} \approx$ 24 meV, similar to the CDW gap recently measured in TaTe$_4$ at 4.2 K \cite{Sun_2020}.

\end{document}